\newcommand{\pt}{\mbox{$p_\mathrm{T}\,$}}
\newcommand{\kt}{\mbox{$k_\mathrm{T}\,$}}
\newcommand{\xe}{\mbox{$x_\mathrm{E}\,$}}

\newcommand{\piz}{\mbox{$\pi^0\,$}}
\newcommand{\iaa}{\mbox{$I_{\mathrm{AA}}\,$}}
\newcommand{\raa}{\mbox{$R_{\mathrm{AA}}\,$}}
\newcommand{\rg}{\mbox{$R_{\gamma}\,$}}

\documentclass[cits]{PoS}

\usepackage{subfigure}

\title{Jet Fragmentation in Medium and Vacuum with the PHENIX Detector}

\ShortTitle{Jet Fragmentation in Medium and Vacuum}

\author{Matthew Nguyen for the PHENIX Collaboration \thanks{The author would like to acknowledge B. Cole and Y.-S. Lai for their useful correspondence with regard to the jet reconstruction results presented in these proceedings. }
	\\
        CERN \\
        E-mail: \email{Matthew.Nguyen@cern.ch}}

\abstract{
One of the most active areas of investigation in relativistic heavy-ion collisions is the study of the jet quenching phenomenon whereby hard partons lose their energy as they traverse the hot, dense matter created in such collisions.    Strong parton energy loss has been observed in central nucleus-nucleus collisions as evidenced by the a large suppression of the yield of high \pt hadrons as compared to the expected yield based on measurements in $p$+$p$ collisions~\cite{1}.  Moreover, measurements of back-to-back correlations of charged hadrons suggest that jet shapes are strongly modified modified by the medium~\cite{2}.    The quantitative interpretation of single and di-hadron measurements is, however, complicated by the fact that the initial parton energy is unknown.    A more informative measurement would be one in which the initial parton energy is known, allowing the determination of the fragmentation function, which may be effectively modified from its vacuum form by the presence of the medium. Two measurements in which the initial parton energy may be estimated are discussed in these proceedings:  jet reconstruction and two-particle correlations using direct photons.  Jet reconstruction in nuclear collisions is challenging due to the large background of soft particles, fluctuations of which give rise to fake jets.   Direct photons can be used to estimate the initial parton energy of the recoil jet without recourse to jet reconstruction algorithms.   However, such studies suffer from a smaller rate and the direct photon signal must be disentangled from a large background of decay photons.   We present jet reconstruction results which use an algorithm suitable for a high multiplicity environment.  We also present results of two-particle correlations using direct photons.  These results are discussed in the context of medium modification to the fragmentation function.
}

\FullConference{XVIII International Workshop on Deep-Inelastic Scattering and Related Subjects, DIS 2010\\
		April 19-23, 2010\\
		Firenze, Italy}

\begin{document}




\section{Jet Reconstruction}

The PHENIX detector contains two opposing central arms each spanning approximately $\pi$/2 radians in azimuth and 0.7 units of pseudo-rapidity around mid-rapidity~\cite{3}.  Each arm contains a charged particle tracking system, a set of ring-imaging Cherenkov counters for electron identification and electromagnetic calorimeters.   Jet are reconstructed from the set of charged particles and photons obtained from these detectors.   The jet energy scale is determined by a GEANT simulation of the detector response.  Significant jet energy corrections arise from neutral hadrons, which are not measured in PHENIX, and losses due to photon conversions. 

Jets are reconstructed using the Gaussian filter algorithm described in~\cite{4}.   The average background level from soft collisions is subtracted, however fluctuations of the background can give rise to fake jets~\cite{5}.  The Gaussian filter uses an angular weighting to differentiate jets from these fluctuations as well as to mitigate the impact of finite acceptance effects.   Further fake rejection is provided by a discriminant which requires that jets be well-collimated.   The effect of the soft background on the jet energy scale is determined by embedding jets from $p$+$p$ events into heavy-ion events~\cite{6}.  

\begin{figure*}[h] 
\subfigure[]{
\includegraphics[width=.49\textwidth]{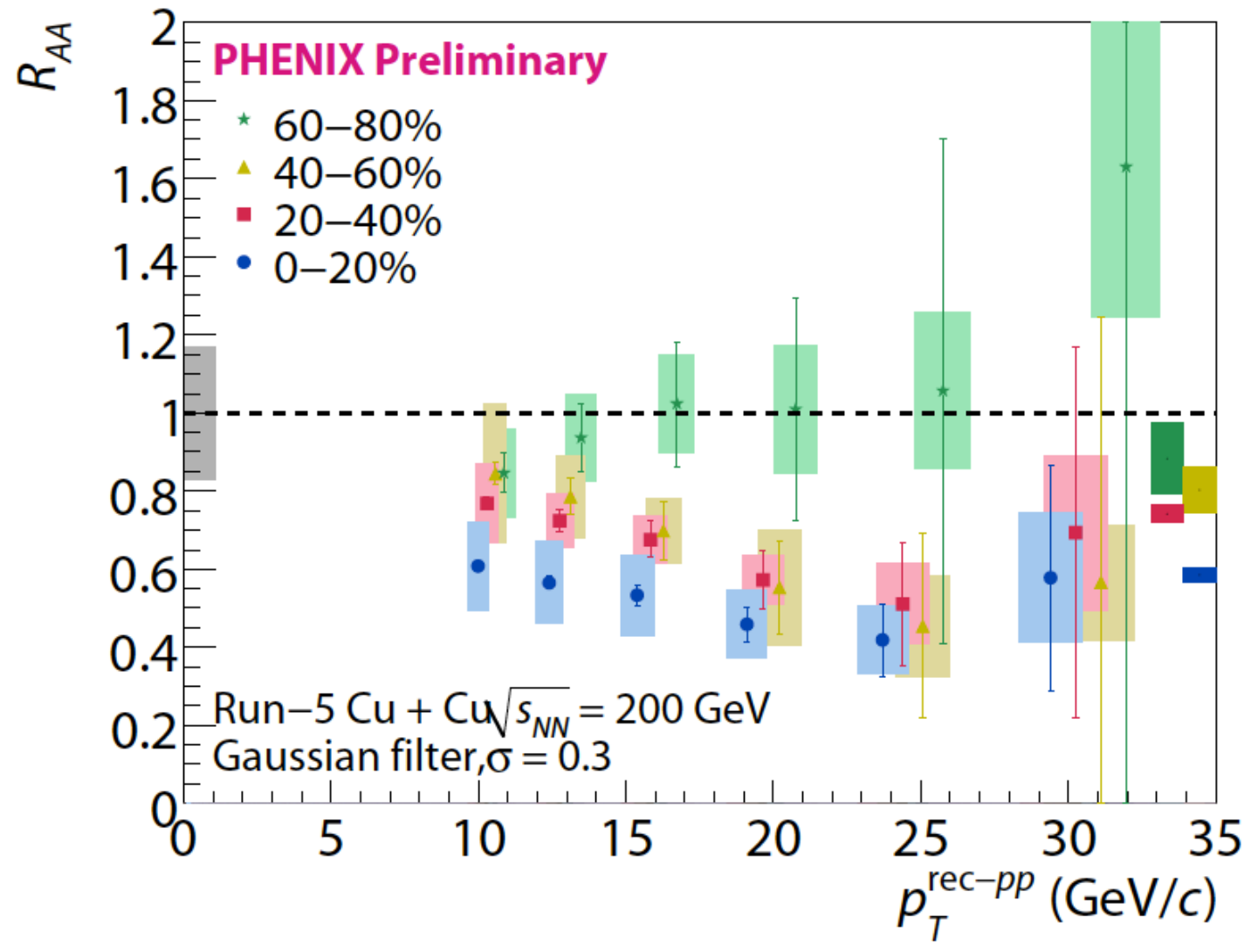}
\label{fig:jet_a}
} 
\subfigure[]{
\includegraphics[width=.51\textwidth]{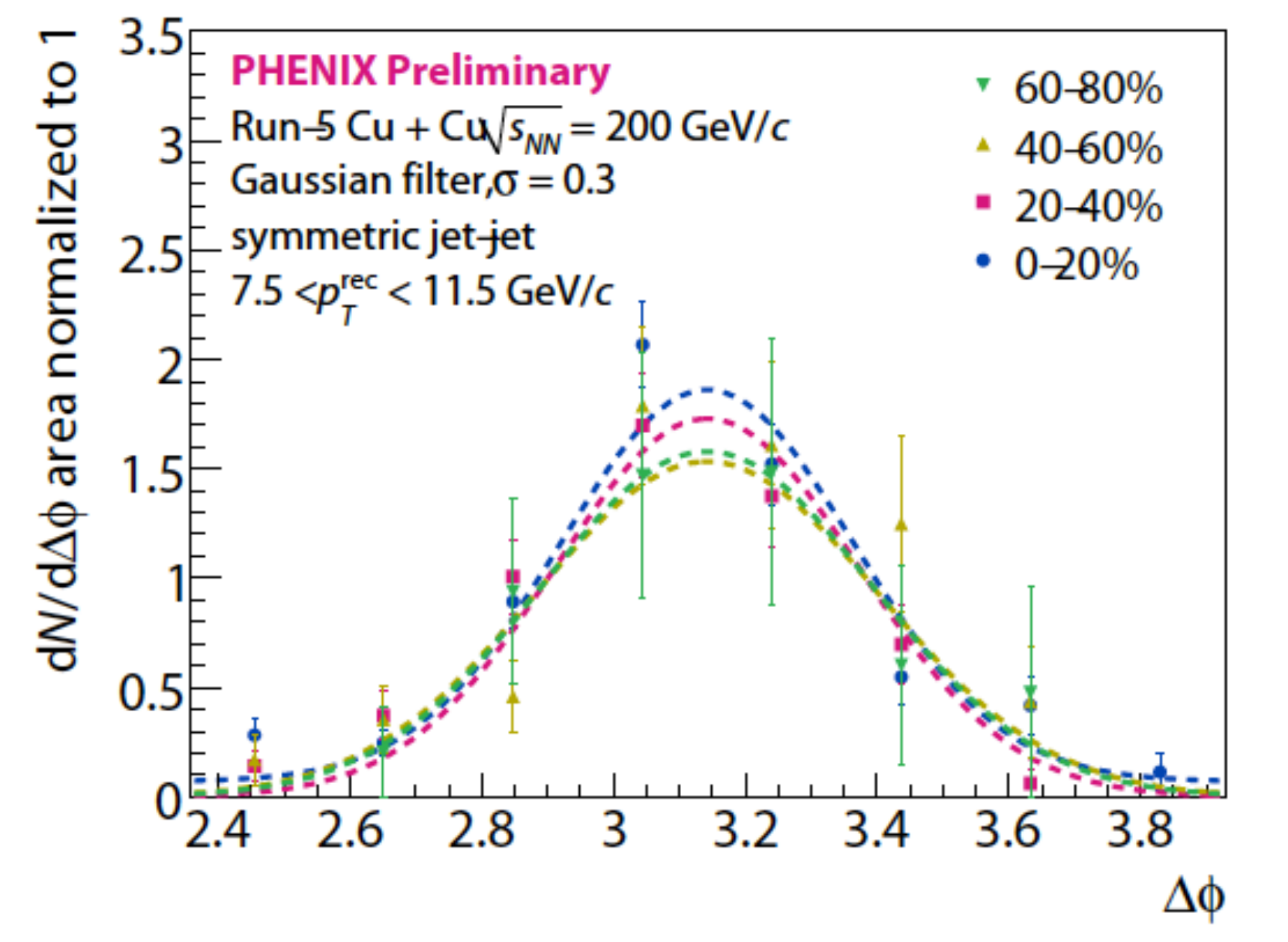}
\label{fig:jet_b}
} 
\caption{(a) Nuclear modification factor, $R_{AA}$, of Gaussian filter jets ($\sigma$ = 0.3) as a function of \pt for four different centrality selections. (b)  Azimuthal jet-jet correlations for several different centrality selections of Cu+Cu collisions using Gaussian filter jets ($\sigma$ = 0.3).
} 
\label{fig:jet} 
\end{figure*}

Figure \ref{fig:jet_a} shows the nuclear modification factor $R_{AA}$ for jets in Cu+Cu collisions for various centrality selections.  The width parameter $\sigma$ of the Gaussian filter (roughly equivalent to the $R$ parameter of other jet algorithms) is set to 0.3.  $R_{AA}$ is the ratio of the yield in nuclear collisions to the yield in $p$+$p$ scaled by the number of binary collisions.   Gaussian filter jets show a significant nuclear modification which is consistent with the value of $R_{AA}$ measured for $\pi^{0}$ in central Cu+Cu collisions.   For peripheral collisions $R_{AA}$ becomes consistent with unity, which would be expected in the absence of nuclear effects.   Azimuthal jet-jet correlations for a symmetric jet pair selection are shown in  \ref{fig:jet_b}.   No significant broadening of the correlation with increasing centrality is evident.   Taken together, these results indicate that jets are sufficiently modified to fall outside the jet definition of Gaussian filter jets with $\sigma = $ 0.3.   Further work is underway to study the fragmentation function from Gaussian filter jets in nuclear collisions.

\clearpage

\section{Direct Photon Correlations}

In contrast to partons, photons have a small cross section for interaction with the dense matter created in central nuclear collisions.   At LO in pQCD the dominant mechanism of direct photon production is the quark-gluon Compton scattering process $q+g \rightarrow q+\gamma$.  Modulo higher order effects, the direct photon can hence be used to determine the initial \pt of the recoil parton before energy loss, since they exactly balance.   Experimentally, the challenge of direct photon measurements is to separate them from the large background of photons produced from hadron decays.

Direct photon correlations with charged hadrons may be obtained by a statistical subtraction of the decay component as follows.   First we measure the per-trigger yield ($Y$) of all photons, which is the number of photon-hadron pairs divided by the number of photons, i.e., $Y_{\rm total} \equiv N_{\gamma-h}/N_{\gamma}$.   This quantity can be decomposed into contributions from direct and decay sources:

\begin{equation}
Y_{\rm total} = \frac{N_{\rm direct}}{N_{\rm total}}Y_{\rm direct} + \frac{N_{\rm decay}}{N_{\rm total}}Y_{\rm decay},
\end{equation}

\noindent where direct is taken to signify any photon not from hadron decay.  We may solve for $Y_{\rm direct}$

\begin{equation}
Y_{\rm direct} = \frac{\rg}{\rg-1} Y_{\rm total} + \frac{1}{\rg-1} Y_{\rm decay},
\end{equation}

\noindent where the direct photon excess, \rg $\equiv N_{\rm total}/N_{\rm decay}$, is known from measurements of the direct photon cross section~\cite{7,8}.  The decay yield $Y_{\rm decay}$ is estimated by measuring correlations of charged hadrons with \piz and $\eta$, which are responsible for more than 95\% of the decay photons.   These decay photon correlations are determined from those of the parent mesons by a simulation which takes into account decay kinematics as well as detector resolution, acceptance and efficiency~\cite{9}.

\begin{figure*}[htb] 
\subfigure[]{
\includegraphics[width=.45\textwidth]{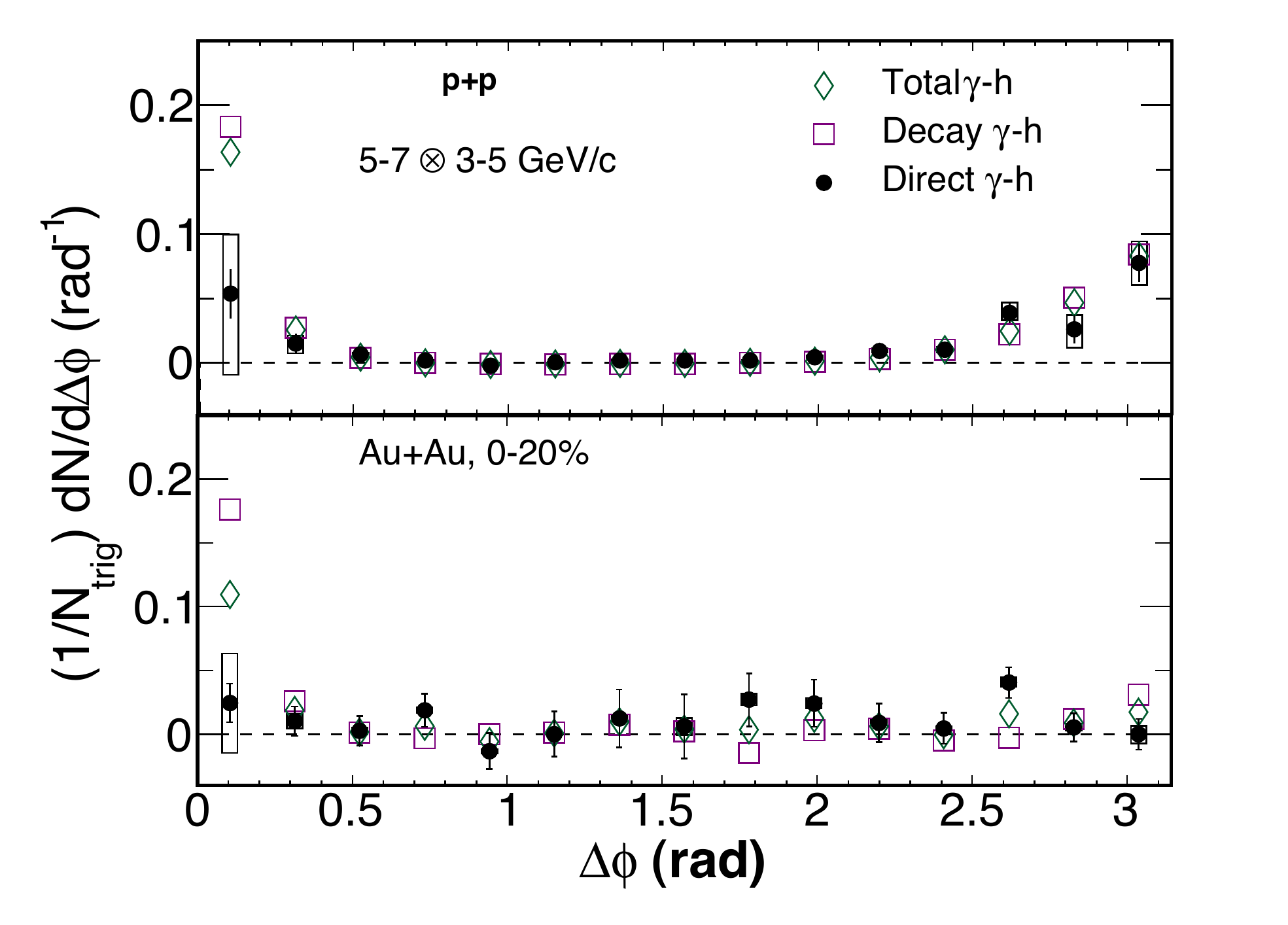}
\label{fig:gjetStat_a}
} 
\subfigure[]{
\includegraphics[width=.54\textwidth]{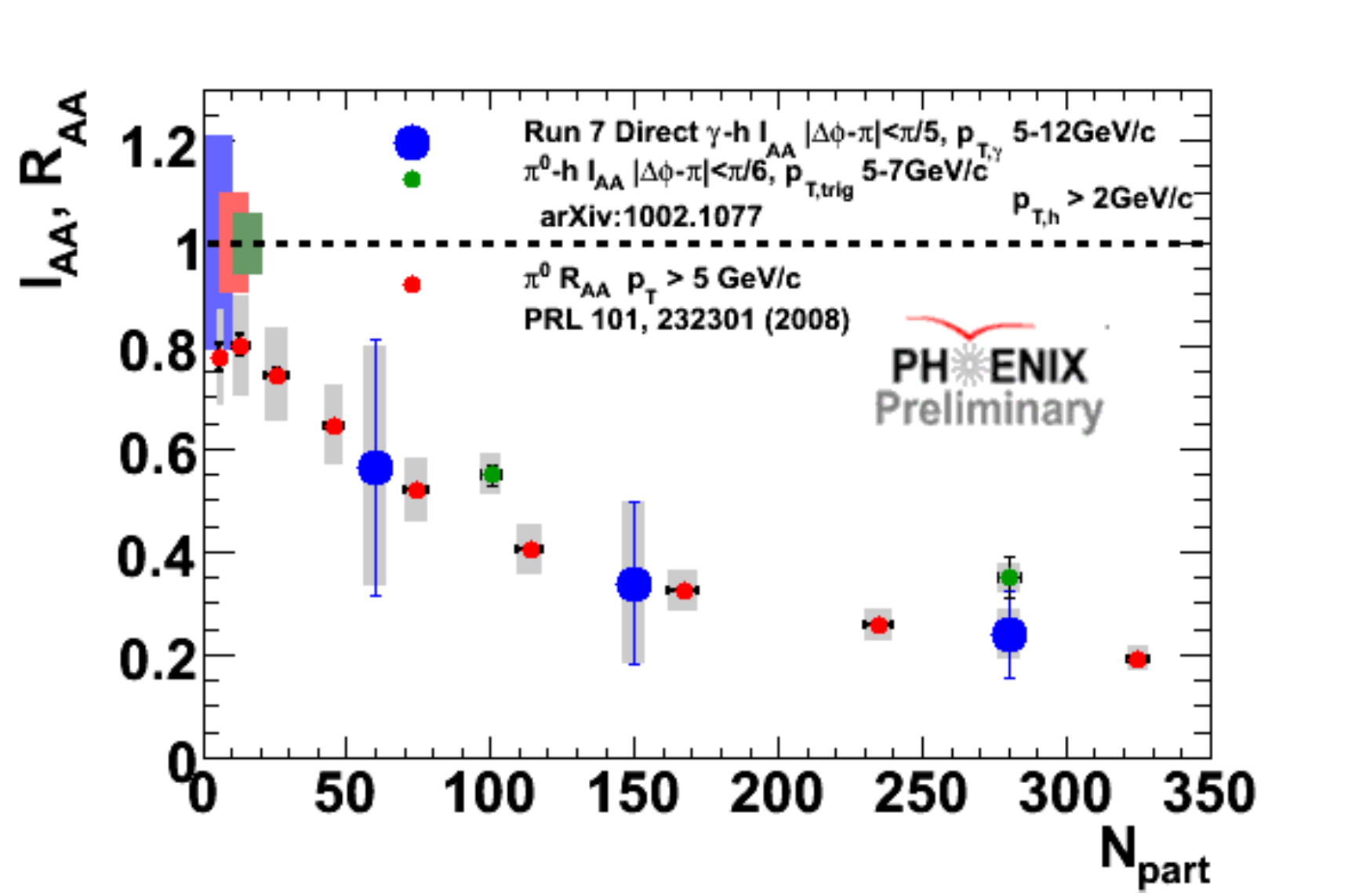}
\label{fig:gjetStat_b}
} 
\caption{(a)  Azimuthal correlations of charged hadrons with all photons, decay photons and direct photons in $p$+$p$ (top panel) and central Au+Au collisions (bottom panel)~\cite{9}.  (b)  Nuclear modification factor for direct photon-hadron correlations in central Au+Au collisions (\iaa) compared to that of \piz-hadron correlations~\cite{10} and of single \piz's (\raa)~\cite{11} as a function of the number of participants.
} 
\label{fig:gjetStat} 
\end{figure*}

 Figure~\ref{fig:gjetStat_a} shows examples of azimuthal correlations, $Y(\Delta \phi)$, for direct photons in $p$+$p$ and central Au+Au collisions, as well as the corresponding total and decay photon correlations used in the statistical subtraction~\cite{9}.   On the near-side $Y_{\rm direct}$ is consistent with zero, as would be expected at LO.  The disappearance of back-to-back correlations in central Au+Au, a signature of parton energy loss well-known from di-hadron correlations, is also evident in $Y_{\rm direct}$.   Fig~\ref{fig:gjetStat_b} shows the nuclear modification factor \iaa, which is the ratio of $Y$ in Au+Au to $Y$ in $p$+$p$, as a function of collision centrality as quantified by the number of participants.   Also shown is \iaa for \piz-hadron correlations~\cite{10} as well as \raa for single \piz's~\cite{11} which both demonstrate a level of modification consistent with the direct photon correlations within sizable uncertainties.   The parton path-length through the medium is expected to vary between the three observables such that for a medium with an extended region of partial transmission, one should observe a different level of modification amongst these observables~\cite{12}.   In contrast, the results suggest that the medium is sufficiently absorbent that the path-length difference is beyond the sensitivity of the current measurements.  

The statistical subtraction of the large decay component results in substantial statistical and systematic uncertainties.   The decay background can be reduced by removing decay photon pairs from \piz and $\eta$ based on their invariant mass and by applying an isolation cut.   Given a finite detector acceptance and efficiency, a residual background remains after the application of these cuts which is subtracted at the statistical level~\cite{13}.  These methods are difficult to apply in high multiplicity nuclear collisions, but are demonstrated in $p$+$p$ in the results which follow.    

\begin{figure*}[htb] 
\subfigure[]{
 \includegraphics[width=.49\textwidth]{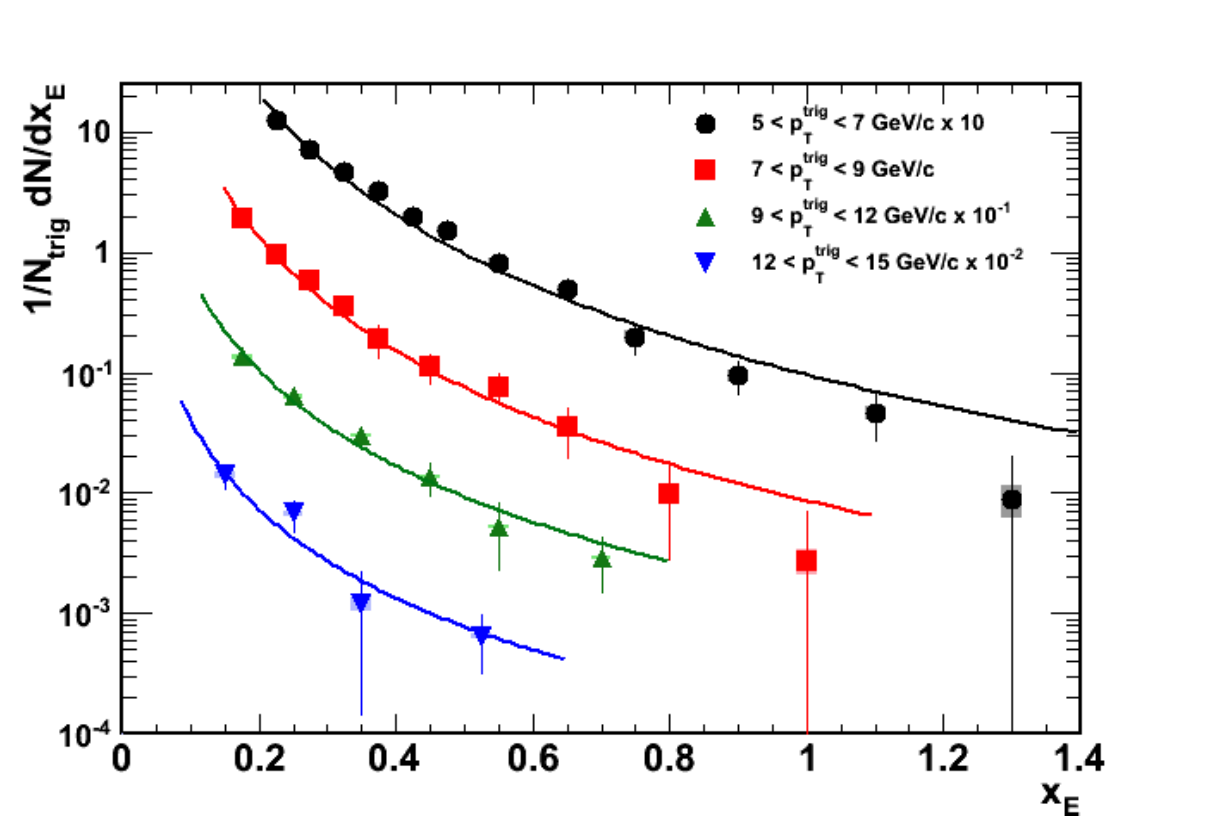} 
 \label{fig:gjetIso_a}
} 
\subfigure[]{
\includegraphics[width=.45\textwidth]{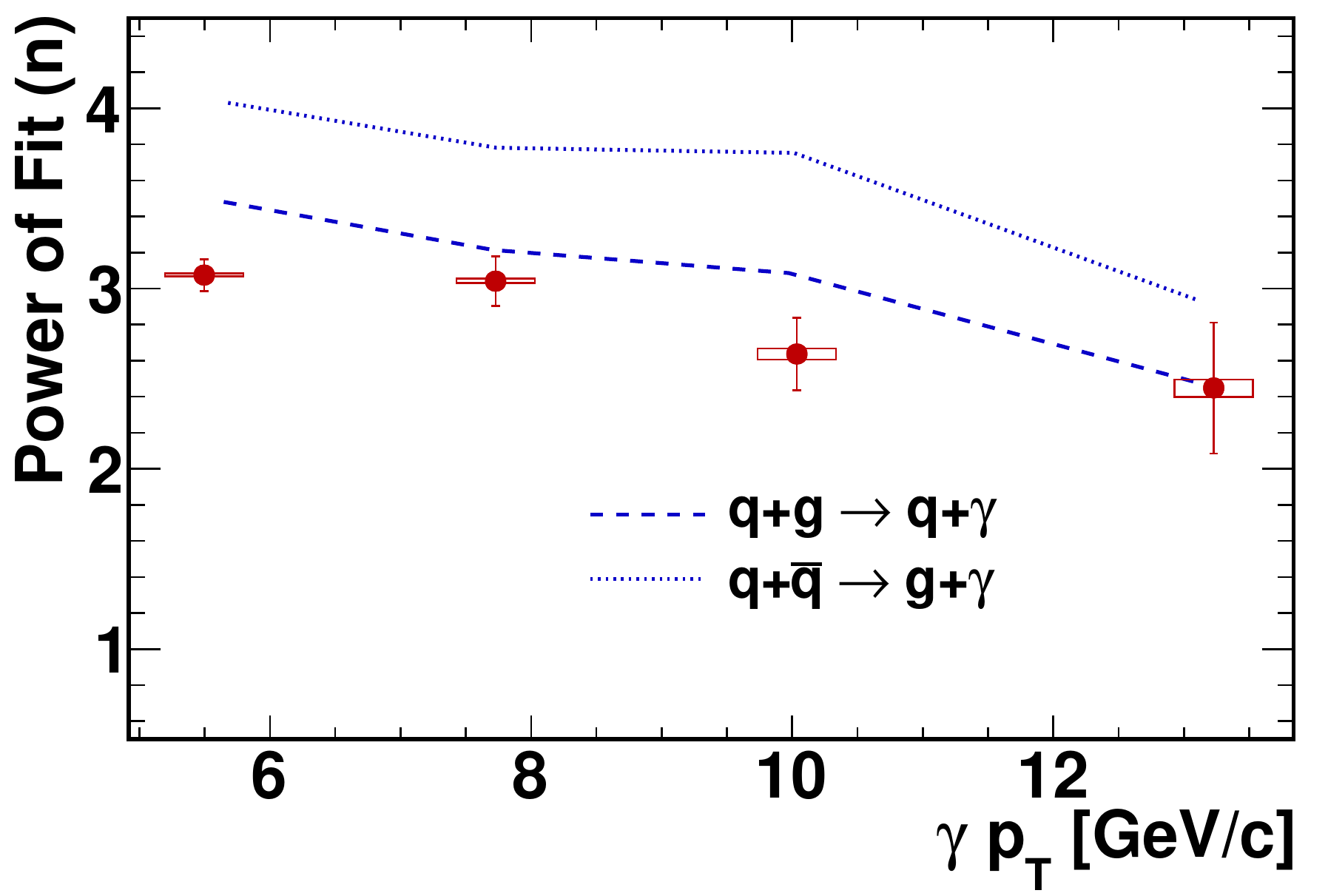} 
 \label{fig:gjetIso_b}
} 
\caption{(a)  Per-trigger yield vs. \xe\ for $\pi^0$ (open symbols) and direct photon triggers (closed symbols) with opposite-side charged hadron partners for several selections of trigger \pt (scaled by factors of 10 for visibility)~\cite{13}.  (b)  Slopes of the power law fits to the direct photon triggered \xe distributions (shown in left panel) compared to a \kt-smeared LO pQCD calculation as described in text~\cite{13}.
} 
\label{fig:gjetIso} 
\end{figure*}

Of particular interest is the fragmentation function of the recoil jet opposite direct photons, which should be effectively modified in-medium~\cite{14}.  The quantity \xe ($\equiv \vec{p}_{\rm T,\gamma} \cdot \vec{p}_{\rm T, h}/|p_{\rm T,\gamma}|^2$) can be used as a proxy for the fragmentation variable $z$, which is a good approximation for $p_{T,\gamma} \approx p_{jet}$. Figure~\ref{fig:gjetIso_a} shows the away-side per-trigger yield as a function of \xe for direct photon triggers in $p$+$p$ collisions.   The values of the parameter $n$ of power-law fits to the \xe distributions (dN/d\xe $\propto \xe^{-n}$) are shown in Fig.~\ref{fig:gjetIso_b}.  To test the sensitivity of the data to the underlying fragmentation function a LO pQCD calculation was performed adding a phenomenological Gaussian \kt smearing whose width was determined from data.   The calculation was performed for both the Compton scattering and annihilation sub-processes, which correspond to quark and gluon fragmentation, respectively.  The data lie closer to the calculation for the Compton sub-process which is known to dominate at the level of about 85-90\%.

\clearpage

\section{Conclusion}

We have presented results on jet reconstruction and direct photon triggered two-particle correlations in $p$+$p$ and nuclear collisions.   Both reconstructed and direct photon-triggered jets were shown to be strongly modified in central nuclear collisions.  Isolated direct photon correlations were shown to be sensitive to the underlying fragmentation function.  Work is underway to measure the medium modifications to the fragmentation function using both observables.

\end{document}